\documentstyle[aps,floats,epsf,twocolumn]{revtex}
\begin{document} \draft

\title{Chiral Symmetry Breaking and Phase Fluctuations in
Cuprate Superconductors:\\
A QED$_3$ Unified Theory of the Pseudogap State}

\author{Z. Te\v{s}anovi\'c$^1$, O. Vafek$^1$ and M. Franz$^2$}
\address{$^1$Department of Physics and Astronomy, Johns Hopkins
University, Baltimore, MD 21218, USA \\
$^2$Department of Physics and Astronomy,
University of British Columbia, Vancouver, BC, Canada V6T 1Z
\\
\rm(\today)}

%
\address{~
\parbox{14cm}{\rm
\medskip
A d-wave superconductor,
its phase coherence progressively destroyed by unbinding of
vortex-antivortex pairs, suffers an
instability related to chiral
symmetry breaking in two-flavor QED$_3$.
The chiral manifold exhibits large degeneracy
spanned by physical states acting as
inherent ``competitors'' of d-wave superconductivity.
Two of these states
are associated with antiferromagnetic insulator
and ``stripe'' phases, known to be stable in the pseudogap
regime of cuprates near half-filling.
The theory also predicts additional, yet unobserved
state: a d+ip phase-incoherent superconductor.
}}
\maketitle


\narrowtext

Ever since the original discovery, the physics of
high temperature superconductors (HTS) has been one of the
key problems in theory of quantum condensed matter.
The most actively pursued approach\cite{anderson}
is to focus on the insulating state of CuO$_2$
planes at half-filling and work one's way along the doping ($x$)
axis to the d-wave superconductor. Alternatively, others
have studied superconducting instabilities of
a nearly antiferromagnetic Fermi liquid (FL)\cite{nafl}.
Both approaches are examples of the traditional
paradigm that ``one should understand
the normal state before one can understand the superconductor'',
the strategy that has met with much success in
conventional, low-$T_c$ superconductors.

Recently, a different route toward the
theory of HTS has been advanced
in Refs. \cite{balents,ftqed}. Cuprates are strongly interacting
systems where traditional approaches might be too forbidding.
Instead, as argued in \cite{ftqed},
one should focus on the superconducting
state itself which appears as the ``least correlated'' among its
neighbors in the HTS phase diagram,
the integrity of its low energy BCS-like quasiparticles protected
by the large d-wave pseudogap.
In this approach one considers the pseudogap regime
as dominated by superconducting phase fluctuations and
seeks to understand the ``normal'' states adjacent to a
superconductor by focusing on
the interaction between quasiparticles and vortex-antivortex excitations.
There is considerable experimental evidence supporting this
viewpoint\cite{emery,corson,ong,fischer,behnia}.
In particular, recent Nernst effect measurements\cite{ong,behnia}
indicate strong vortex fluctuations at temperatures
comparable to the pseudogap ($T \sim T^*$) and
far above the true $T_c$. The effective
low energy theory of these interactions was argued to be quantum
electrodynamics in (2+1) dimensions (QED$_3$)\cite{ftqed}.

The success of this new approach hinges on its ability,
by using the d-wave
superconducting state as its starting point,
to reconstruct the general features of the HTS phase diagram.
Amongst these, none is more prominent than the Neel antiferromagnetic
insulator very near half-filling.  In this Letter we first show
that a d-wave superconductor whose phase coherence
has been destroyed by unbinding of quantum vortex-antivortex
pairs indeed becomes an antiferromagnet. This confirms  
important result of Herbut\cite{herbut}.
The antiferromagnetism
arises naturally through an inherent dynamical instability
of QED$_3$, known as the spontaneous chiral symmetry
breaking (CSB)\cite{appelquist}, and most typically takes the form
of an incommensurate spin-density-wave (SDW), whose periodicity
is tied to the Fermi surface. 
Furthermore, we next show that numerous other
states, most notably a d+ip and a d+is
{\em phase-incoherent} superconductors (dipSC, disSC) and ``stripes'',
i.e. superpositions of 1D charge-density-waves (CDW) and
{\em phase-incoherent} superconducting-density-waves (SCDW),
as well as continuous chiral rotations among them, are all energetically
close and competitive with antiferromagnetism due to their
equal membership in the chiral manifold of two-flavor ($N=2$) QED$_3$.
This large chiral manifold of nearly degenerate states
plays the key role in our theory as the culprit behind
the complexity of the HTS phase diagram.

The above results place tight restrictions on this
phase diagram and provide means to unify the phenomenology of cuprates
within a single, systematically calculable
``QED$_3$ Unified Theory'' (QUT). Any {\em microscopic} description
of cuprates, as long as it leads to the large d-wave pairing pseudogap
with $T^* \gg T_c\to 0$, will conform to the general results
of QUT. In particular, all the physical states in natural energetic proximity
to a d-wave superconductor are the ones inhabiting the above
chiral manifold. Under the umbrella of the pseudogap,
the energetics and various properties of such states are
explicitly calculable from the chirally {\em symmetric}
QED$_3$ theory of Ref. \cite{ftqed} which plays the
role in the pseudogap state similar to that of the FL
theory in conventional metals.

We now provide the substance behind the above assertions.
Our starting point is the QED$_3$ Lagrangian
\begin{equation}
{\cal L}_{\rm QED}=
\bar\psi_n c_{\mu,n}\gamma_\mu D_\mu \psi_n
+{\cal L}_0[a_\mu] + (\cdots),
\label{qed}
\end{equation}
shown in Ref. \cite{ftqed} to desribe the low energy effective theory for
fermions in a d-wave superconductor interacting with dynamically fluctuating
vortex excitations of the Cooper pair field.
Here $\psi_\alpha^{\dag} = \bar\psi_\alpha\gamma_0=
(\eta_\alpha^\dag, \eta_{\bar\alpha}^\dag)$
are the four-component Dirac spinors  with
$\eta_\alpha^\dag = \frac{1}{\sqrt 2}\Psi^\dag_\alpha
({\bf 1}+i\sigma_1)$,
$\eta_{\bar\alpha}^\dag = \frac{1}{\sqrt 2}\Psi^\dag_{\bar\alpha}\sigma_2
({\bf 1}+i\sigma_1)$,
and $\Psi^\dag_\alpha = (\psi^\dag_{\uparrow\alpha},\psi_{\downarrow\alpha})$.
Fermion fields $\psi_{\sigma\alpha}({\bf r},\tau)$ describe `topological
fermions' of the theory and are related to the original nodal
fermions $c_{\sigma\alpha}({\bf r},\tau)$ via
the singular gauge transformation detailed in Refs. \cite{ftqed,ft}.
Index $n$ labels  $(1,\bar{1})$ and $(2,\bar{2})$ {\em pairs}
of nodes while $\alpha$ labels individual nodes,
$\mu = \tau, x, y (\equiv 0, 1, 2)$.
$D_\mu = \partial_\mu +ia_\mu$ is a covariant derivative,  $c_{\tau,n} =1$,
$c_{x,1} = c_{y,2}=v_F$,
$c_{x,2} = c_{y,1}=v_\Delta $. The gamma matrices are defined as
$\gamma_0=\sigma_3\otimes\sigma_3$,
$\gamma_1=-\sigma_3\otimes\sigma_1$,
$\gamma_2=-\sigma_3\otimes\sigma_2$,
and satisfy $\{\gamma_\mu,\gamma_\nu\}=2\delta_{\mu\nu}$. The Berry
gauge field $a_\mu$ encodes the
topological frustration of nodal fermions generated
by fluctuating quantum vortex-antivortex pairs and
${\cal L}_0$ is its bare action. The loss of superconducting
phase coherence caused by unbinding of vortex pairs is
heralded in (\ref{qed}) by $a_\mu$ becoming massless:
\begin{equation}
{\cal L}_0 \to
\frac{1}{2e^2_\tau} (\partial\times a)^2_\tau
+\sum_i \frac{1}{2e^2_i}(\partial\times a)^2_i~;
\label{2+1d}
\end{equation}
here $e^2_\tau , e^2_i (i=x,y)$, as well as the
velocities $v_{F(\Delta)}$,  are functions of doping $x$ and $T$.
Along with residual interactions between nodal fermions,
denoted by the ellipsis in (\ref{qed}), these parameters of QUT
arise from some more microscopic description and
will be discussed shortly.

First, however, we focus on the general
properties of (\ref{qed}).
The Berry gauge field $a_\mu$
plays a special role in the above
expression. If we set $a_\mu =0$, all the remaining interactions
among nodal fermions are short-ranged, including those
arising from the integration over the Doppler gauge
field $v_\mu$,\cite{ftqed} and irrelevant in the RG
sense. They can impact the low energy physics
only through symmetry breaking and frequently first-order transitions.
Consequently, if  $a_\mu$ were absent or massive, like in
the superconducting state, the effective theory
of the pseudogap state would be that of free, massless Dirac
fermions. In contrast, $a_\mu$ is relevant in the massless
(non-superconducting) state and it generates non-trivial
long range interactions among quasiparticles.
The effective theory of the pseudogap state is
QED$_3$ and the low energy physics is controlled by the
{\em interacting} infrared fixed point of its chiral {\em symmetric}
(massless fermion) phase. This is the ``algebraic'' Fermi liquid normal state
discussed in Ref. \cite{ftqed}.

${\cal L}_{\rm QED}$ (\ref{qed}) possesses the following
peculiar continuous symmetry: borrowing from ordinary quantum
electrodynamics in (3+1) dimensions (QED$_4$),
we know that there exist two additional
gamma matrices, $\gamma_3 = \sigma_1\otimes {\bf 1}$ and
$\gamma_5 = i\sigma_2\otimes {\bf 1}$ that anticommute
with {\em all} $\gamma_\mu$.
We can define a global U(2) symmetry for each pair of nodes,
with generators ${\bf 1}\otimes {\bf 1}$, $\gamma_3$,
$-i\gamma_5$ and $\frac{1}{2}[\gamma_3,\gamma_5]$, which leaves
${\cal L}_{\rm QED}$ invariant. In QED$_3$ this symmetry can be broken
by two ``mass'' terms, $m_{\rm ch}{\bar\psi}_n\psi_n$
and $m_{\rm PT}{\bar\psi}_n \frac{1}{2}[\gamma_3,\gamma_5]\psi_n$.
Spontaneous symmetry breaking in QED$_3$ as a mechanism for dynamical mass
generation has been extensively studied in the field theory literature
\cite{appelquist,parity,maris}. It has been established that
while $m_{\rm PT}$ is never spontaneously generated\cite{parity},
the chiral mass $m_{\rm ch}$ is generated if number
of fermion species $N$ is less than a critical value $N_c$. It is found that
$N_c\sim 3$ for {\em isotropic} QED$_3$\cite{maris,cohen}, but as we
shall discuss shortly, anisotropy and irrelevant couplings present in
Lagrangian (\ref{qed}) can change the value of $N_c$.

\begin{figure}[t]
\epsfxsize=6.5cm
\hspace{1cm}
\hfil\epsfbox{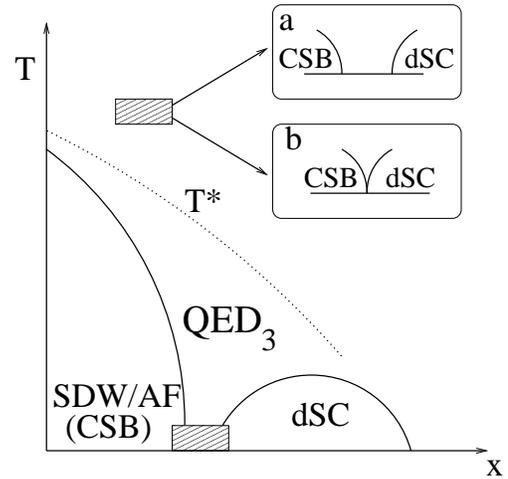}\hfill
\vspace{.25cm}
\caption[]{Schematic phase diagram of a cuprate superconductor
in QUT. Depending on the value of $N_c$ (see text), either
the superconductor
is followed by a {\em symmetric} phase of QED$_3$ which then undergoes
a quantum CSB transition at some lower doping (panel a), or
there is a direct transition from the superconducting phase
to the $m_{\rm ch}\not= 0$ phase of QED$_3$ (panel b).
The label SDW/AF indicates
the dominance of the antiferromagnetic ground state as $x\to 0$.
}
\label{fig1}
\end{figure}

Let us now assume that we are in the part of the phase diagram (Fig. 1)
characterized by the parameters such that $N_c>2$: CSB occurs and
the mass term $m_{\rm ch}{\bar\psi}_n\psi_n$ is generated. We wish to
determine what is the nature of this chiral
instability in terms of the original electron operators.  To make this
apparent, let us consider a general chiral rotation
$\psi_n\to U_{\rm ch}^{(n)}\psi_n$ with
$U_{\rm ch}^{(n)}=\exp (i\theta_{3n}\gamma_3 + \theta_{5n}\gamma_5)$.
Within our representation of Dirac spinors (\ref{qed}),
the $m_{\rm ch}{\bar\psi}_n\psi_n$ mass term takes the
following form:
$$m_{\rm ch}\cos (2\Omega_n)\bigl [\eta^\dag_\alpha\sigma_3\eta_\alpha
- \eta^\dag_{\bar\alpha}\sigma_3\eta_{\bar\alpha}\bigr ] +$$
\begin{equation}
+ m_{\rm ch}\sin(2\Omega_n)\frac{\theta_{5n} +
i\theta_{3n}}{\Omega_n}\eta^\dag_\alpha\sigma_3\eta_{\bar\alpha}
+ {\rm h.c.}~~,
\label{chiral}
\end{equation}
where $\Omega_n = \sqrt{\theta_{3n}^2 + \theta_{5n}^2}$.
$m_{\rm ch}$ acts as an order parameter for the bilinear combinations of
topological fermions appearing in (\ref{chiral}). In the
symmetric phase of QED$_3$ ($m_{\rm ch}=0$) the expectation values of
such bilinears vanish, while they become finite,
$\langle{\bar\psi}_n\psi_n\rangle \not= 0$, in the
broken symmetry phase.

The chiral manifold (\ref{chiral}) is spanned by the ``basis''
of three symmetry breaking
states. When reexpressed in terms of the original nodal fermions
$c_{\sigma\alpha}({\bf r},\tau)$,
two of these involve pairing in the particle-hole (p-h)
channel -- a cosine and a sine spin-density-wave (SDW):
$$
\langle c^\dag_{\uparrow\alpha}
c_{\uparrow\bar\alpha} -
c^\dag_{\downarrow\alpha}c_{\downarrow\bar\alpha}\rangle +
{\rm h.c.}~~~~~~~~~({\rm cos~SDW})
$$
\begin{equation}
i\langle  c^\dag_{\uparrow\alpha}
c_{\uparrow\bar\alpha} -
c^\dag_{\uparrow\bar\alpha}
c_{\uparrow\alpha}\rangle + (\uparrow\to
\downarrow)~~~~~~~({\rm sin~SDW})
\label{sdw}
\end{equation}
and are obtained from Eq. (\ref{chiral}) by setting
$\Omega_n$ equal to $\pi/4$ or $3\pi/4$. Rotations
within the chiral manifold (\ref{chiral}) at fixed
$\Omega_n$ correspond to the sliding
modes of SDW.

\begin{figure}[t]
\epsfxsize=4.5cm
\hspace{1.5cm}
\hfil\epsfbox{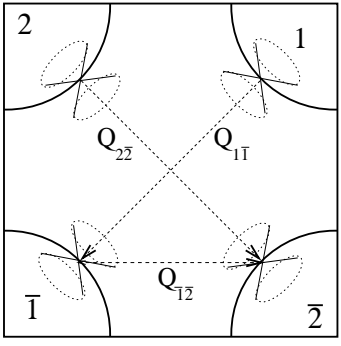}\hfill
\vspace{.5cm}
\caption[]{The ``Fermi surface'' of cuprates, with the positions
of nodes in the d-wave pseudogap. The wavectors ${\bf Q}_{1\bar 1},
{\bf Q}_{2\bar 2}, {\bf Q}_{\bar 1\bar 2}$, etc. are discussed
in the text.
}
\label{fig2}
\end{figure}

A simple
physical picture emerges here: we started from a d-wave
superconducting phase, our parent state. As one moves
closer to half-filling and
true phase coherence is lost, strong vortex-antivortex
pair fluctuations, acting under the protective umbrella of
a d-wave particle-particle (p-p)
pseudogap, spontaneously induce formation of particle-hole ``pairs''
at finite wavevectors $\pm  {\bf Q}_{1\bar 1}$ and
$\pm {\bf Q}_{2\bar 2}$, spanning
the Fermi surface from node $\alpha$ to $\bar\alpha$ (Fig. 2).
The glue that binds these p-h ``pairs'' and plays the
role of ``phonons'' in this pairing analogy is provided by the
Berry gauge field $a_\mu$.
Such ``fermion duality'' is a natural consequence of the
QED$_3$ theory (\ref{qed}). Remarkably, we find the
antiferromagnetic insulator being spontaneously generated
in form of the incommensurate SDW.
As we get very near half-filling and
${\bf Q}_{1\bar 1}, {\bf Q}_{2\bar 2}$ approach
$(\pm\pi,\pm\pi)$, SDW acquires the
most favored state status within the chiral
manifold -- this is the consequence of umklapp processes
which increase its condensation energy without it being
offset by either the anisotropy or
a poorly screened Coulomb interaction which
plagues its CDW competitors to be introduced shortly.
It seems therefore reasonable to argue
that this SDW must be considered the progenitor of the
Neel-Mott-Hubbard insulating
antiferromagnet at half-filling. Thus, QED$_3$ theory
(\ref{qed}) explains the origin
of antiferromagnetic order in terms of strong vortex-antivortex
fluctuations in the parent d-wave superconductor. It does
so naturally, through its inherent and
well-established chiral symmetry breaking instability\cite{appelquist}.

The chiral manifold (\ref{chiral}) contains also a third state,
a p-p pairing state corresponding to
$\Omega_n =0$ or $\pi/2$  and
best characterized as a d+ip
phase-incoherent superconductor:
\begin{equation}
i\langle\psi_{\uparrow\alpha}\psi_{\downarrow\alpha}-
\psi_{\uparrow\bar\alpha}\psi_{\downarrow\bar\alpha}\rangle +
{\rm h.c.}~~~~~~~~~~~~({\rm dipSC})~.
\label{dipsc}
\end{equation}
We have written dipSC in terms of
topological fermions $\psi_{\sigma\alpha}({\bf r},\tau)$
since use of the original fermions
leads to more complicated expression which involves the backflow of
vortex-antivortex excitations described by gauge fields
$a_\mu$ and $v_\mu$ (such backflow terms do not arise
in the p-h channel).
This state breaks parity but
preserves time reversal, translational invariance and
superconducting U(1) symmetries. To our knowledge, such
state has not been proposed as a part of any of the major
theories of HTS. It is an intriguing question
whether this d+ip phase-incoherent superconductor
can be the actual ground state at some
dopings in some of the cuprates. Its energetics does not
suffer from long range Coulomb problems but it is clearly
inferior to the SDW very close to half-filling since, being
spatially uniform, it receives no help from umklapp.
Observation of such a state in underdoped cuprates would
provide strong evidence for the validity of the
physical picture proposed in this Letter.

Until now, we have discussed the CSB pattern only within
individual pairs of nodes, (1,$\bar 1$) and (2,$\bar 2$).
What happens if we allow for
chiral rotations that mix nodes 1 and $\bar 2$ or
1 and 2? A whole new plethora of states becomes possible,
with chiral manifold enlarged to include a superposition of
{\em one-dimensional} p-h and p-p states,
an incommensurate CDW accompanied by
a non-uniform phase-incoherent superconductor (SCDW)
at wavevectors $\pm {\bf Q}_{12}$
and $\pm {\bf Q}_{\bar 2\bar 1}$ (Fig. 2):
$$\frac{1}{\sqrt{2}}\langle c^\dag_{\uparrow 1}c_{\uparrow 2}
+ c^\dag_{\uparrow\bar 2}c_{\uparrow\bar 1}+ {\rm h.c.}\rangle
+ (\uparrow\rightarrow\downarrow)~~~{\rm (CDW)}$$
\begin{equation}
\frac{1}{\sqrt{2}}\langle \psi_{\uparrow 1}\psi_{\downarrow 2}
+ \psi_{\uparrow\bar 2}\psi_{\downarrow\bar 1}+ {\rm h.c.}\rangle
+ (\uparrow\leftrightarrow\downarrow)~~{\rm (SCDW)}
\label{stripes}
\end{equation}
These same states, rotated by $\pi/2$, are replicated at wavevectors
$\pm {\bf Q}_{1\bar 2}$
and $\pm {\bf Q}_{2\bar 1}$ (Fig. 2). In a fluctuating
d$_{x^2-y^2}$ superconductor these CDWs and SCDWs run along
the $x$ and $y$ axes and are naturally identified as the
``stripes'' of QUT.
Note, however, these are not the only
one-dimensional states in QUT -- among the states
in the chiral manifold (\ref{chiral}) are also
``diagonal stripes'', the combination of
a SDW (\ref{sdw}) along $\pm {\bf Q}_{1\bar 1}$ and
a dipSC (\ref{dipsc}) which opens the mass gap only at nodes
$(2,\bar 2)$, or vice versa.
Furthermore, a phase-incoherent d+is superconductor (disSC)
is also present within the chiral enlarged manifold, since
it results in alternating signs for
different nodes with equal number of positive
and negative ``masses'' for the two-component nodal fermions:
\begin{equation}
i\langle\psi_{\uparrow 1}\psi_{\downarrow 1}+
\psi_{\uparrow\bar 1}\psi_{\downarrow\bar 1}+
{\rm h.c.}\rangle + (1\rightarrow 2)~~~~({\rm disSC})~.
\label{dissc}
\end{equation}
In contrast, in a d+id
phase incoherent superconductor these ``masses'' have
the same sign for all the nodes producing a maximal
breaking of the PT symmetry\cite{parity}. Consequently, a d+id
phase-incoherent superconductor is not spontanously induced
within our QED$_3$ theory.

In the {\em isotropic} ($v_F=v_\Delta$) $N=2$
QED$_3$  all these additional states
plus arbitrary {\em chiral} rotations
among them
are completely equivalent to those dicussed previously.
It is here where we confront the problem of intrinsic
anisotropy in Eq. (\ref{qed}). Such anisotropy cannot be
rescaled out and manifestly breaks the U(2)$\times$U(2)
degeneracy of the full $N=2$ chiral
manifold down to two separate
U(1)$\times$U(1) (\ref{chiral}) chiral groups
discussed previously. This is reflected in the
general increase in energy of the states from the enlarged
chiral manifold. For example, the anisotropy
raises the energy of our
``stripe'' states (\ref{stripes}) relative to
those of SDW, dipSC or ``diagonal stripes''.
However, when the long
range Coulomb interactions and coupling to the lattice
are included in the problem, as they are in real materials,
it is conceivable that the ``stripes'' would return in some
form, either as a ground state or a long-lived metastable
state at some intermediate doping.
disSC is also adversely affected by anisotropy but to
a lesser extent
and might remain competitive with SDW, dipSC
and ``diagonal stripes''\cite{qutlong}.
This state breaks time reversal symmetry but preserves parity and
the discussion concerning dipSC below Eq. (\ref{dipsc})
applies to disSC equally well.

How do we use these general results on CSB in QUT to address
the specifics of cuprate phase diagram? To this end, we
need some effective combination of phenomenology and
more microscopic descriptions to determine
the parameters $v_F$, $v_\Delta$, $e_\tau$, $e_i$
and residual interactions $(\cdots)$ appearing in
${\cal L}_{\rm QED}$ (\ref{qed}).
The main task is to determine what is the
sequence of states within QUT that form stable phases
as the doping decreases toward half-filling under $T^*$
in Fig. 1.
While this is an extensive project whose detailed results will
be reported elsewhere\cite{qutlong}, we
outline here some of the general features.
First, within the superconducting
state $e_\tau, e_i\to 0$ and $a_\mu$ becomes massive thus
denying the CSB mechanism its main dynamical agent. We
therefore expect that the superconductor is in the symmetric
phase and its nodal fermions form well-defined excitations\cite{ftqed}.
As we move to the left in Fig. 1, the phase order is
suppressed and $e_\tau, e_i$ become finite, reflecting the
unbinding of vortex-antivortex excitations\cite{ftqed}. For all practical
purposes, this is precisely what the experiments imply.
Now, the key question is whether the QED$_3$ (\ref{qed})
remains in its symmetric phase or whether it immediately
undergoes the CSB transition and generates finite gap
($m_{\rm ch}\not= 0$).

One important factor in the above problem is the dependence of $N_c$ on the
Dirac cone anisotropy $\alpha_D=v_F/v_\Delta$. Intuitively one would expect
that $N_c$ decreases (see also Ref. \cite{herbut})
with increasing anisotropy because the phase space
for the interactions that ultimately drive the CSB transition
is reduced as the overlap between the two pairs of Dirac cones with opposite
anisotropies diminishes. 
In the superconducting state close to optimal doping
it is known that anisotropy is fairly large, $\alpha_D\simeq 10-20$ 
\cite{chiao1}. We expect that in this case $N_c(\alpha_D) < N=2$; near
optimum doping we are far from the chiral mass generation.
On underdoping the pseudogap size $\Delta$
increases which implies decreasing $\alpha_D$ and increasing $N_c$.
Eventually, with sufficient underdoping, $N_c$ exceeds the value of $N=2$ and
a quantum  phase transition occurs into the state with broken chiral
symmetry as illustrated in Fig. 1.

However, the anisotropy is not the only factor which can
influence the value of $N_c$. Short range interactions, while
perturbatively irrelevant, effectively increase $N_c$ if
stronger than some critical value\cite{gusynin}. Such interactions,
typically in the form of short 
range three-current terms\cite{balents}, arise
in more microscopic models used
to derive ${\cal L}_{\rm QED}$\cite{qutlong} and are prominent among
the residual terms denoted by ellipsis in (\ref{qed}). Their
strength generically increases as $x\to 0$.
These residual interactions play a dual role in QUT.
First, they can conspire with the anisotropy to
produce the situation depicted
in panel (b) of Fig. 1, where the CSB takes place as soon
as the phase coherence is lost.
Second, once the chiral symmetry has been broken, the residual
interactions further break the symmetry within the chiral
manifold (\ref{chiral}) and play a role in selecting
the true ground state. A detailed analysis of the CSB patterns
will be reported separately\cite{qutlong}.

The authors are indebted to 
Prof. I. F. Herbut for helpful discussions. 
This work was
supported in part by NSF 
grant DMR00-94981 (ZT and OV) and by NSERC (MF).


\begin{references}
\bibitem{anderson} P. W. Anderson, {\em The theory of
superconductivity in the high-T$_c$ cuprates},
Princeton University Press (1997).
\bibitem{nafl} P. Monthoux and  D. Pines, \prb {\bf 47}, 6069 (1993).
\bibitem{balents} L. Balents, M. P. A. Fisher and C. Nayak, \prb {\bf 60},
1654 (1999).
\bibitem{ftqed} M. Franz and Z. Te\v sanovi\' c, cond-mat/0012445.
\bibitem{emery} V. J. Emery and S. A. Kivelson, Nature {\bf 374},
434 (1995).
\bibitem{corson} J. Corson {\em et al.}, Nature {\bf 398}, 221 (1999).
\bibitem{ong} Z. A. Xu {\em et al.}, Nature {\bf 406}, 486 (2000);
Y. Wang {\em et al.}, cond-mat/0108242.
\bibitem{fischer} B. W. Hoogenboom {\em et al.}, \prb  {\bf 62}, 9179 (2000).
\bibitem{behnia} C. Capan {\em et al.},  cond-mat/0108277.
\bibitem{herbut} I. F. Herbut, cond-mat/0110188.
\bibitem{appelquist} R. D. Pisarski, \prd {\bf 29}, 2423 (1984);
T. W. Appelquist {\em et al.}, \prd {\bf 33},
3704 (1986); M. R. Pennington and D. Walsh, Phys. Lett. B {\bf 253}, 246
(1991); I. J. R. Aitchison and N. E. Mavromatos, \prb {\bf 53}
9321 (1996).
\bibitem{ft} M. Franz and Z. Te\v sanovi\' c, \prl {\bf 84}, 554 (2000);
O. Vafek {\em et al.}, \prb {\bf63}, 134509 (2001).
\bibitem{parity} T. Appelquist {\em et al.}, \prd {\bf 33}, 3774 (1986).
\bibitem{maris} P. Maris, \prd {\bf 54}, 4049 (1996).
\bibitem{cohen} See, however, T. Appelquist, A. G. Cohen and M. Schmaltz,
\prd {\bf 60}, 045003 (1999).
\bibitem{qutlong} O. Vafek, M. Franz and Z. Te\v sanovi\' c, in preparation.
\bibitem{chiao1} M. Chiao {\em et al.}, \prb {\bf 62}, 3554 (2000).
\bibitem{gusynin} V. Gusynin, A. Hams and
M. Reenders, \prd {\bf 63}, 045025 (2001).
\end{references}
\end{document}